\pdfoutput=1 
\documentclass[preprint2]{aastex6}
\usepackage[T1]{fontenc}
\usepackage{lmodern}
\usepackage{graphicx}
\usepackage{amsmath}
\usepackage{amsfonts}
\usepackage{amssymb}

\newcommand{\figfactor}{1.0}
\newcommand{\figfactortwo}{1.0}

\newcommand{\Dcal}{{\mathcal{D}}}
\newcommand*{\diff}{\mathop{}\!\mathrm{d}}

\begin{document}

\title{Blazar Variability From Turbulence in Jets Launched by Magnetically 
Arrested Accretion Flows}
\shorttitle{}
\author{Michael O' Riordan\altaffilmark{1}$^\star$, 
    Asaf Pe'er\altaffilmark{1}, 
    Jonathan C. McKinney\altaffilmark{2}}
\shortauthors{O' Riordan, Pe'er, \& McKinney}

\altaffiltext{1}{Physics Department, University College Cork, Cork, Ireland}
\altaffiltext{2}{Department of Physics and Joint Space-Science Institute,
    University of Maryland, College Park, MD 20742, USA}

\email{$^\star$michael\_oriordan@umail.ucc.ie}

\begin{abstract}

Blazars show variability on timescales ranging from minutes to 
years, the former being comparable to and in some cases even shorter than the 
light-crossing time of the central black hole.
The observed $\gamma$-ray lightcurves can be described by a power-law power 
density 
spectrum (PDS), with a similar index for both BL Lacs and
flat-spectrum radio quasars.
We show that this variability can be produced by turbulence in relativistic 
jets launched by magnetically arrested accretion flows (MADs).
We perform radiative transport calculations on the 
turbulent, highly-magnetized jet launching region of a MAD with a rapidly 
rotating supermassive black hole. 
The resulting synchrotron and synchrotron self-Compton emission, 
originating from close to the black hole horizon, is highly variable.
This variability is characterized by PDS which is remarkably similar to the 
observed power-law spectrum at frequencies less than a few per day.
Furthermore, turbulence in the jet launching region naturally produces 
fluctuations in the plasma on scales much smaller than the horizon radius.
We speculate that similar turbulent processes, operating in the jet at large
radii (and therefore high bulk Lorentz factor), are responsible for 
blazar variability over many decades in frequency, including on minute 
timescales.

\end{abstract}
\maketitle


\section{Introduction}
\label{sec:intro}
It is generally accepted that blazars are active galaxies with relativistic jets 
aligned close to our line of sight \citep[e.g.,][]{UP95}. 
Broadly speaking, blazars can be classified into two main categories: 
low-luminosity BL Lacs and high-luminosity flat-spectrum radio quasars (FSRQs),
forming the so-called ``blazar sequence'' \citep{Fossati+98,Ghisellini+17}.
These objects are variable at all observed wavelengths from
radio to $\gamma$-rays \citep[e.g.,][]{Ulrich+97}.
In particular, the $\gamma$-ray lightcurves show variability on timescales 
ranging from minutes to years. 
This variability can be characterized as having a power density 
spectrum (PDS) of power-law shape, 
spanning the entire observed frequency range.
The power law index is remarkably similar for both BL Lacs and FSRQs.
\citet{Abdo+10f} report a PDS slope of $1.4\pm 0.1$ for 9 bright FSRQs, as well 
as slopes of $1.7\pm 0.3$ and $1.5\pm 0.2$ for 6 BL Lacs and 13 faint FSRQs,
respectively.
Recently, \citet{Ackermann+16} report a slope of $1.24\pm 0.15$ in the case 
of the FSRQ 3C 279, and \citet{Goyal+17} report a slope of $1.1\pm 0.2$ for 
the BL Lac object PKS 0735+178. 
\citet{Max-Moerbeck+14} provide estimates of the $\gamma$-ray PDS slopes for 
29 blazars, using simulated lightcurves to properly account for noise processes.
Their findings are largely consistent with the results of \citet{Abdo+10f}.
A similar analysis was performed by \citet{Ramakrishnan+15} for a sample of 55 
blazars from the first 5 years of \emph{Fermi}/LAT data.
They find average slopes of $1.3$ and $1.1$ obtained from 35 
FSRQs and 12 BL Lacs, respectively. These slopes are somewhat smaller than those 
reported by \citet{Abdo+10f}, however, the slopes reported for 3C 279 are
consistent in both cases.

The radio emission, resolved at large radii, clearly originates in jets.
Apparent superluminal motion provides compelling evidence that the
radio
emitting matter propagates with relativistic velocities \citep[e.g.,][]{UP95}.
Since the $\gamma$-rays are unresolved, however, the source of high-energy 
emission 
is uncertain and could potentially be located much closer to the central 
supermassive black 
hole \citep[e.g.,][]{PM12}. 
Generally speaking, the observed short timescale variability implies a 
compact emission region. 
While such a compact region close to the black hole might be responsible for
variability in low-luminosity systems, 
high-luminosity
systems (such as 3C 279) require a significant Lorentz factor to overcome the
pair opacity which would otherwise prevent $\gamma$-rays 
from escaping to infinity
\citep[see e.g., Appendix~\ref{sec:pairs} and][]{DG95}.
The location of the $\gamma$-ray emission zone remains a topic of active 
research \citep[e.g.,][for a recent review]{MS16}.

The shortest variability timescales 
are comparable to, and in some cases shorter than, 
the light-crossing time of the black hole 
\citep{Aharonian+07,Albert+07,Aleksic+11,Ackermann+16}.
For example, \citet{Aharonian+07} observed variability on a timescale of 
$t_\text{var}\approx 200s$ during a flare in PKS 2155--304.
This is more than an order magnitude shorter than the corresponding 
light-crossing time, inferred by the empirical relation between galactic bulge 
luminosity and black hole mass \citep{Bettoni+03,Aharonian+07}.
The mechanism responsible for this fast variability is poorly understood.
Popular models include the ``jets in a jet'' model
\citep{GUB09}, in which magnetic reconnection in highly-magnetized regions of 
the jet accelerates compact blobs of plasma to relativistic velocities in the 
bulk jet frame;
the magnetospheric acceleration model 
\citep{LR11}, in which charged particles are accelerated by unscreened electric 
fields in a charge starved vacuum gap of the black hole magnetosphere;
the relativistic turbulence model \citep{NP12},
in which magnetohydrodynamic turbulence 
in the jet produces compact blobs on scales smaller than the horizon radius, 
similar to those in the ``jets in a jet'' scenario; and the jet-star interaction
model \citep{Barkov+12} in which stars cross the jet close to the black hole.

In this work, we argue in favour of the proposal by \citet{NP12}, 
namely that turbulence in the 
relativistic jet can produce the observed variability properties.
To support this claim, we investigate the variability of high-energy 
radiation from a 
magnetically arrested accretion flow \citep[MAD;][]{N+03},
which efficiently launches jets via the Blandford-Znajek (BZ)
mechanism \citep{BZ77,TNM11,MTB12}.
Importantly, computational limitations force us to restrict our analysis to the 
inner $r\lesssim 200 r_g$ of the MAD, where $r_g = GM/c^2$ is the 
gravitational radius.
\citet{O'Riordan+16b} showed that, for rapidly rotating black holes,
observers see deep into the hot dense, highly-magnetized plasma in the inner
parts of MADs. In what follows, we will refer to this region as the 
``jet launching region''.
Since the radiated power in our model is dominated by the turbulent plasma 
of the jet launching region close to the 
horizon, we expect the resulting radiation to be variable on timescales 
comparable to the light-crossing time $t_g = r_g / c$.

We find that $\gamma$-ray lightcurves from the jet launching region
can be described by a PDS which is 
remarkably similar to that observed in both BL Lacs and FSRQs.
Furthermore, the optical synchrotron emission can also be described by the same 
PDS, despite having a different origin.
The large inferred pair opacity in high-luminosity FSRQs prevents radiation 
originating in the jet
launching zone from directly contributing to the observed variability. 
However, assuming that the turbulent properties of 
the launching region persist at large radii (and therefore large bulk Lorentz 
factor $\Gamma$), we argue that it is plausible that the observed variability 
properties result from turbulence in the jet.

In Section~\ref{sec:model} we briefly describe our model and assumptions.
In Section~\ref{sec:results} we show the resulting PDS and compare
with observations. 
In Section~\ref{sec:discussion} we discuss our findings, with 
emphasis on the limitations of our model and suggestions for future work in this
area.
We use units where $G=c=1$ and therefore $r_g=t_g=M$, 
however we occasionally reintroduce factors of $c$ for clarity.

\section{Model}
\label{sec:model}
We simulate a MAD accretion flow using the
fully 3D general-relativistic magnetohydrodynamic 
(GRMHD) code \texttt{HARM} \citep{Gammie+03}. 
Our model is based on run A0.99N100 from \citet{MTB12}, restarted at
$t=15000\,M$ with a very high temporal resolution of $\Delta t = 0.1M$,
and a spatial resolution of 
$N_r\times N_\theta\times N_\phi=288\times 128\times128$.
This is the highest spatial resolution available for a 3D GRMHD simulation of a 
MAD accretion flow.
As described in \citet{TNM11,MTB12}, the grid concentrates resolution in the
equatorial disk at small radii and in the jet at large radii.
Close to the black hole, the radial grid is logarithmically spaced with 
$\Delta r \approx 0.03\,M$ at the horizon, increasing to $\Delta r \approx M$ at 
$r = 30\,M$.
For our purposes, we consider a black hole spin of $a=0.99$ and
limit our analysis to times $t\geq 15000\,M$, 
well after the simulation settled into a quasi-steady MAD state.
As is discussed in \citet{O'Riordan+16a,O'Riordan+16b}, we remove the numerical 
density floor 
material from the centre of the funnel region before performing our radiative 
transport post-processing calculation. 
Although this material is required to maintain numerical stability during the 
GRMHD simulation,
it can become artificially dense and hot and so might distort the resulting 
spectra.
Therefore, when calculating the spectra, we simply remove the floor material, 
focussing our attention on the
self-consistent disk and funnel wall regions.
We do this by setting the density to zero
in regions where the ratio of magnetic and rest mass energy densities 
becomes too large; $b^2/\rho>\zeta$. Here,
$b^2=b^\mu b_\mu$, $b^\mu$ is the magnetic 4-field, $\rho$ is 
the rest mass density of the fluid.
At the horizon, we choose $\zeta=20$, and linearly interpolate to $\zeta=10$ at 
$r=10\,M$. Beyond this, we choose a constant value of $\zeta=10$. This ensures 
that we don't remove any of the
self-consistent, highly-magnetized fluid close to the horizon when discarding
the floor material.
 
We calculate the properties of the resulting radiation field
using a code based on \texttt{grmonty} 
\citep{Dolence+09}. 
We include synchrotron emission, self-absorption, and Compton scattering from 
relativistic, thermal electrons. 
We choose a constant proton-to-electron temperature
ratio of $T_p/T_e=30$, 
which is consistent with recent radiative GRMHD 
simulations of active galactic nuclei \citep{Sadowski+17b}.
As discussed in \citet{Ressler+15,Foucart+16,Sadowski+17a},
electrons in the jet and disk
are probably subject to different heating and cooling mechanisms due to 
differences in density and magnetization. This would likely cause $T_p/T_e$ to
vary across these regions.
In our case, the emission is strongly dominated by a highly-magnetized, 
compact region close to the 
black hole and we find that varying $T_p/T_e$ independently in the jet and disk
has little effect on our main results.
Furthermore, choosing a different ratio of $T_p/T_e=3$ everywhere doesn't
affect the resulting temporal behaviour.
The absence of non-thermal electrons, which could be important in a more 
realistic description of the high-energy $\gamma$-ray lightcurves, is a 
limitation of the current model; we will address this in a future work.

For analysing the variability, we choose snapshots of the GRMHD data 
corresponding to time steps $\Delta t$ ranging
from $0.1M$ to $10M$. 
This allows us study variability on a wide range of timescales, while still
producing lightcurves which are evenly sampled in time.
We also vary the total number of photons tracked during our radiative transport
calculation. The number of photons tracked per time step ranges from $10^5$, for  
calculations over long timescales, to $10^7$ per time step for the shortest 
timescales.
Importantly,
we use a ``fast-light'' approximation in which 
each snapshot is treated independently. This allows us to parallelize our 
radiative transport calculation over both photons and time steps,
significantly improving both performance and computational simplicity.
Since the emission from our model originates in the compact jet launching region, 
this approximation is valid over a wide range of timescales.
The disadvantage of this approach, however, is that it sets practical 
limitations on the shortest timescales that we can reliably probe. 
The ``fast-light'' assumption effectively sets the speed of 
light to infinity, which fails to be valid for timescales shorter than the 
light-crossing time of the emission region.
Therefore, our current code is not sensitive to very fast variability on
timescales shorter than $\sim M$.
Furthermore, the smallest structures in the GRMHD data are smoothed out over 
$\sim 10$ cells and so, on the shortest timescales, our analysis is also limited 
by the grid resolution.

\newpage
\section{Results}
\label{sec:results}
\begin{figure}
    \centering
    \includegraphics[width=\figfactor\linewidth]{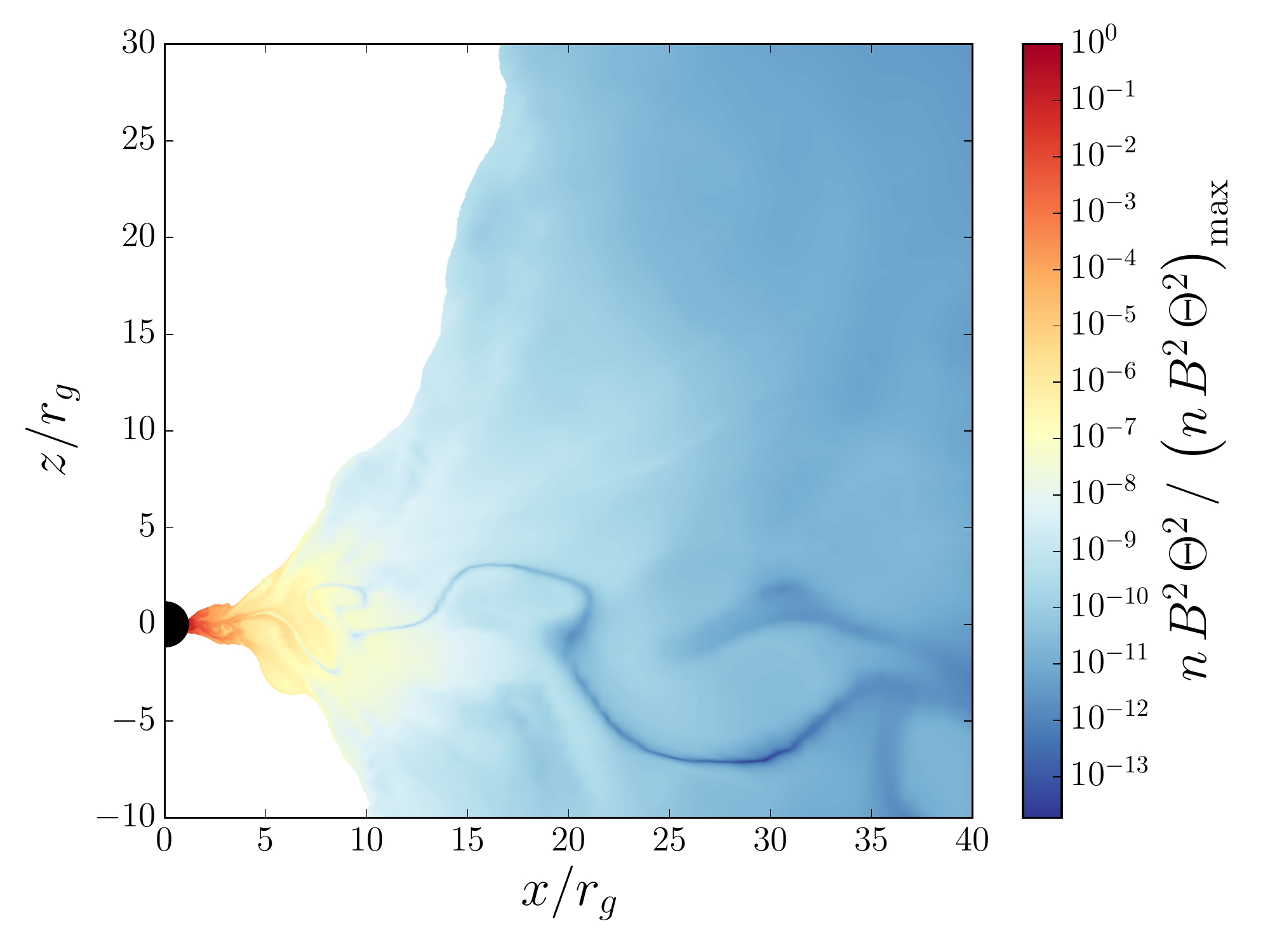}
    \caption{Snapshot of the GRMHD model showing the quantity $nB^2\Theta^2$,
        which is proportional to the comoving synchrotron power per unit
        volume. The floor material has been excised from the funnel region.
        Both the synchrotron power and the synchrotron self-Compton component
        are strongly dominated by the inner $\sim 5\,M$.
}
    \label{fig:nB2T2}
\end{figure}
\begin{figure}
    \centering
    \includegraphics[width=\figfactortwo\linewidth]{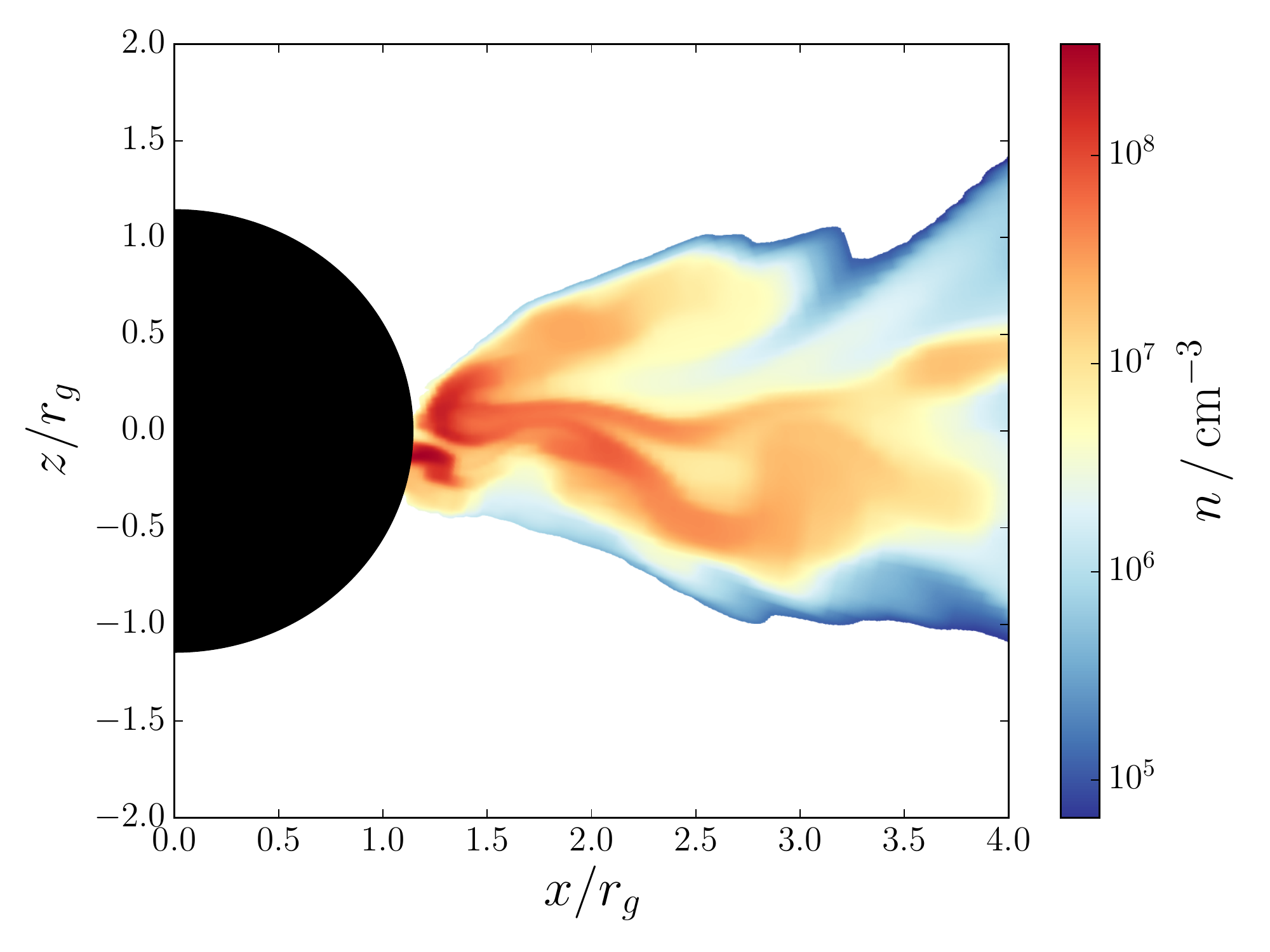}
    \includegraphics[width=\figfactortwo\linewidth]{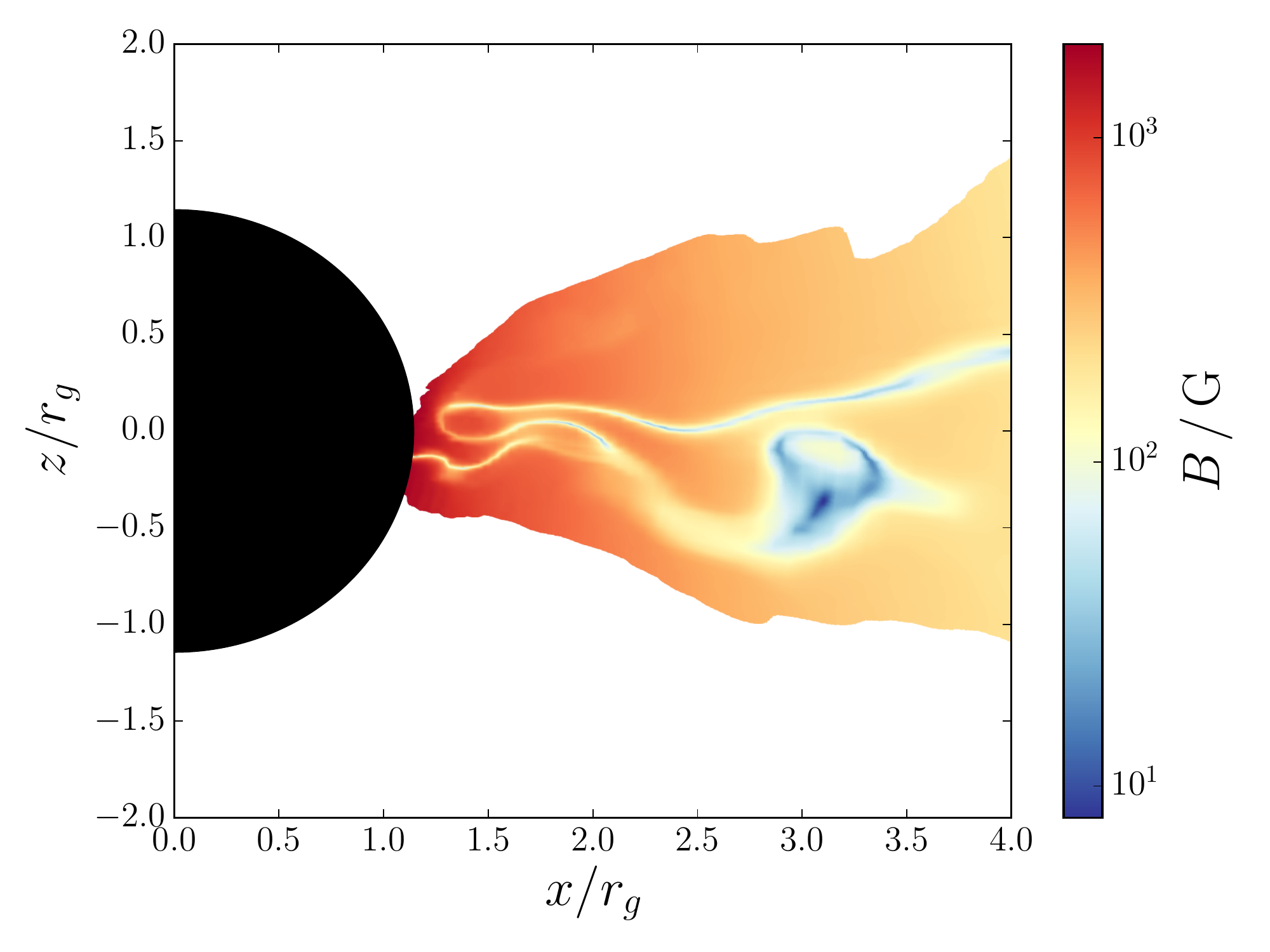}
    \caption{Snapshot of the GRMHD model at $t=15700\,M$. The top panel shows
    the comoving electron number density and the bottom panel shows the comoving 
    magnetic field
    strength. The white regions correspond to floor material which has been 
    removed. Clearly, there are fluctuations in both $n$ and $B$ on scales
    smaller than $r_g$.
}
    \label{fig:n_and_B}
\end{figure}

\begin{figure}
    \centering
    \includegraphics[width=\figfactor\linewidth]{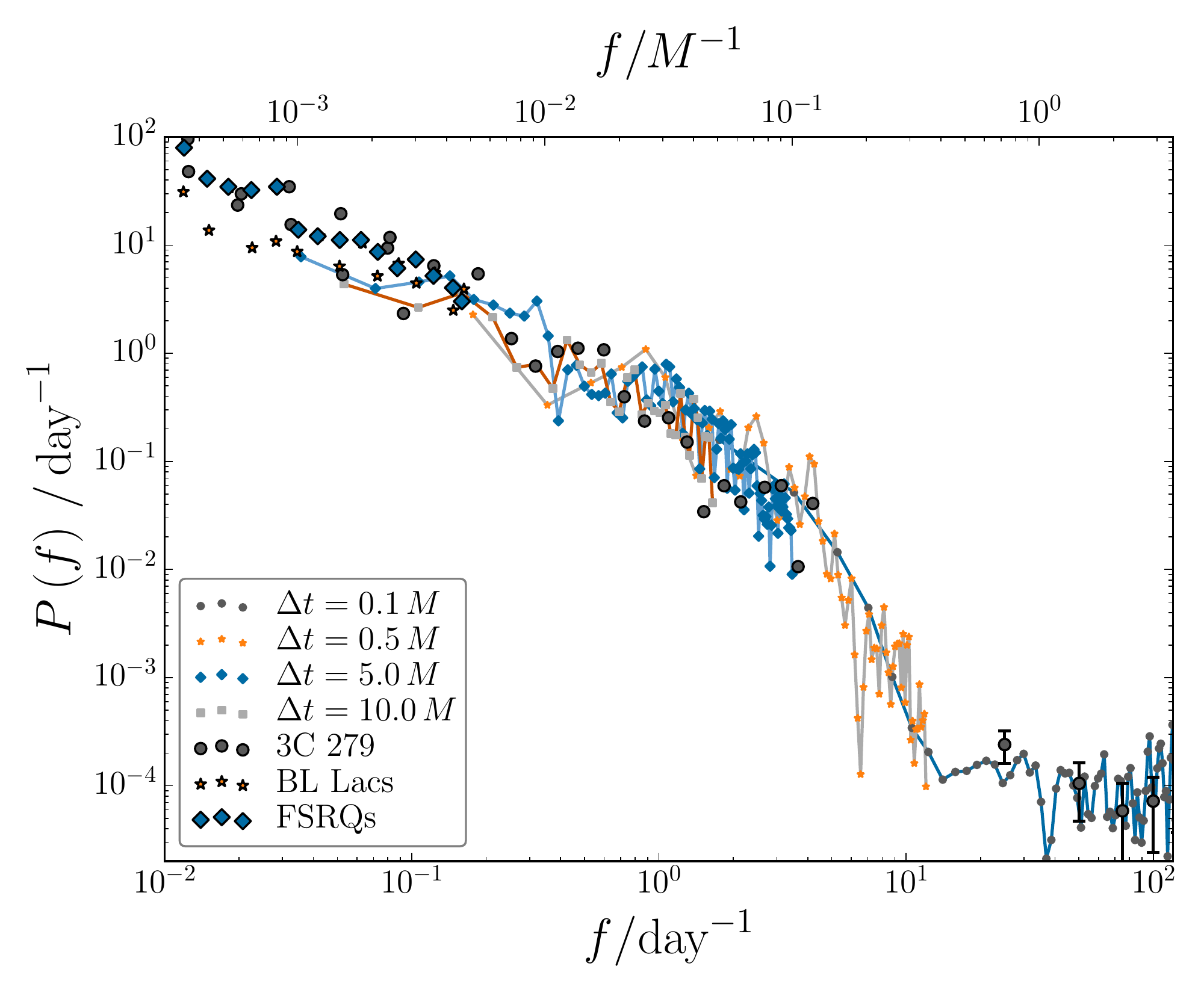}
    \caption{$\gamma$-ray PDS corresponding to four lightcurves frequency-integrated 
        between $\nu_\text{min} = 10^{19}\,\text{Hz}$ and 
        $\nu_\text{max} = 10^{24}\,\text{Hz}$. 
        The time steps show the sampling of each lightcurve.
        The circles correspond to observational data 
        from 3C 279 \citep{Ackermann+16}, 
        while the stars and diamonds correspond to data averaged over 
        BL Lacs and FSRQs \citep{Abdo+10f}. 
        The observational data corresponds to energies $>100\,\text{MeV}$.
        Both BL Lacs and FSRQs show similar power-law 
        behaviour, and the simulated PDS agrees reasonably well with both of 
        these.
        There is a clear cut-off in the spectrum above 
        $\sim 5\,\text{day}^{-1}$. The white noise part of the spectrum above
        $10\,\text{day}^{-1}$ is due to poor photon statistics.
        Note that the top axis is scaled to a black hole mass of 
        $M=5\times 10^{8}\,M_\odot$ for comparison with the data for 3C 279,
        while the averaged data from \citet{Abdo+10f} corresponds to systems
        with different masses.
        The normalization on the y-axis is arbitrary, and simply chosen for
        comparing the shape of our simulated PDS to the observations.
}
    \label{fig:gamma_ray_psd}
\end{figure}

For a thermal electron distribution, the synchrotron power radiated in the 
comoving frame scales as
$P_\text{syn}\sim\int\diff V\, nB^2\Theta^2$,
where $\diff V=\sqrt{-g}\diff x^r \diff x^\theta \diff x^\phi$, 
$g=\det\left(g_{\mu\nu}\right)$ is the metric determinant, 
$n$ is the electron number density, $B$ is the magnetic field,
$\Theta=kT/mc^2$ is the electron temperature, and all fluid quantities are 
measured in the fluid frame.
In Figure~\ref{fig:nB2T2} we show a snapshot of the fluid data at $t=15700\,M$.
The white region aligned with the spin axis corresponds to the numerical floor 
material that has been removed from the funnel.
We plot the quantity $nB^2\Theta^2$, which is proportional to the comoving
synchrotron power per unit volume. Clearly, the synchrotron emission in this 
model is strongly dominated by the inner $\sim 5\,M$. Similarly, the 
synchrotron self-Compton component originates in a compact region close to the
black hole. 
Therefore, the variability in the resulting lightcurves is dominated by the 
turbulent, highly magnetized plasma in the inner parts of the MAD. 
Since the emission is dominated by fluid at small radii,
there is no significant Lorentz boosting of the radiation. 

In the top panel of Figure~\ref{fig:n_and_B} we show a snapshot of the comoving 
electron number density close to the black hole horizon. 
In the bottom panel we show the corresponding comoving magnetic field strength.
The flow is turbulent, and there are significant fluctuations in the fluid 
properties extending down to spatial scales much smaller than $M$.
Notably, there are structures in the density apparent on scales 
$\lesssim 0.5\,M$, and structures in $B$ apparent on scales of
$\lesssim 0.1\,M$ due to a polarity change of the poloidal magnetic field.
The resolution of our numerical model means that we can only resolve 
such small scale structures close to the horizon. However, we expect that 
similar sub-$M$ features could be produced by 
turbulent processes operating at large radii.
In particular, if such inhomogeneities are produced at large distances in a 
highly-magnetized, relativistic jet, it is plausible that they could contribute 
significantly to variability on very short timescales.

The PDS for real signal $h(t)$ is
$P(f) = 2\left|\hat{h}(f)\right|^2$,
where $\hat{h}(f)$ is the Fourier transform of $h(t)$ and the frequency range
is $0\leq f < \infty$ \citep[see e.g.,][]{Press1986}.
To numerically estimate the PDS, we follow a similar procedure to that
described in \citet{Wellons+14}.
To reduce noise, we divide each lightcurve into two segments and 
average the resulting power spectra. Furthermore, to suppress leakage between 
frequency bins,we apply a Hann window to each segment \citep[e.g,][]{Press1986}.
In practise, we use the implementation provided by the ``welch'' function in 
the SciPy Python module \citep{scipy}.
In Figure~\ref{fig:gamma_ray_psd} we show the PDS calculated from four
$\gamma$-ray lightcurves 
(frequency integrated between $10^{19}$ and
$10^{24}\,\text{Hz}$) with time steps ranging from $0.1\,M$ to $10\,M$. 
The resulting PDS is not particularly sensitive to the sampling of the 
lightcurves, with good agreement in regions of overlap.
These $\gamma$-rays result from inverse-Compton scattering 
by hot electrons in the jet launching region.
For comparison with observations, the bottom axis shows the frequency in units
of $\text{day}^{-1}$, while
the top axis shows the frequency in units of inverse light-crossing time,
normalized to a black hole mass of $M=5\times 10^{8}\,M_\odot$.
We have also included data for 3C 279 from \citet{Ackermann+16} as well as the 
average PDS for 6 BL Lacs and 9 FSRQs from \citet{Abdo+10f}. 
Note that the scaling of the top axis is appropriate for comparison with the 
3C 279 data, while the \citet{Abdo+10f} data is averaged over different masses.
Remarkably, our simulated PDS is largely consistent with the data from both 
BL Lacs and FSRQs (including 3C 279) below 
$f\sim\,\text{few}\,\,\text{day}^{-1}$.
This result is suggestive of a connection between blazar variability and 
turbulence in the jet launching region of MADs.
Note that since we are primarily interested in the scaling with $f$, 
the normalization of the y-axis is arbitrary, and simply chosen to ease 
comparison between the shape of our simulated PDS and the data. 
Also, since the overall magnitude of the PDS is probably sensitive 
to the prescription for treating the electron temperature, a detailed 
investigation of the normalization is beyond the scope of the this work.

\begin{figure}
    \centering
    \includegraphics[width=\figfactor\linewidth]{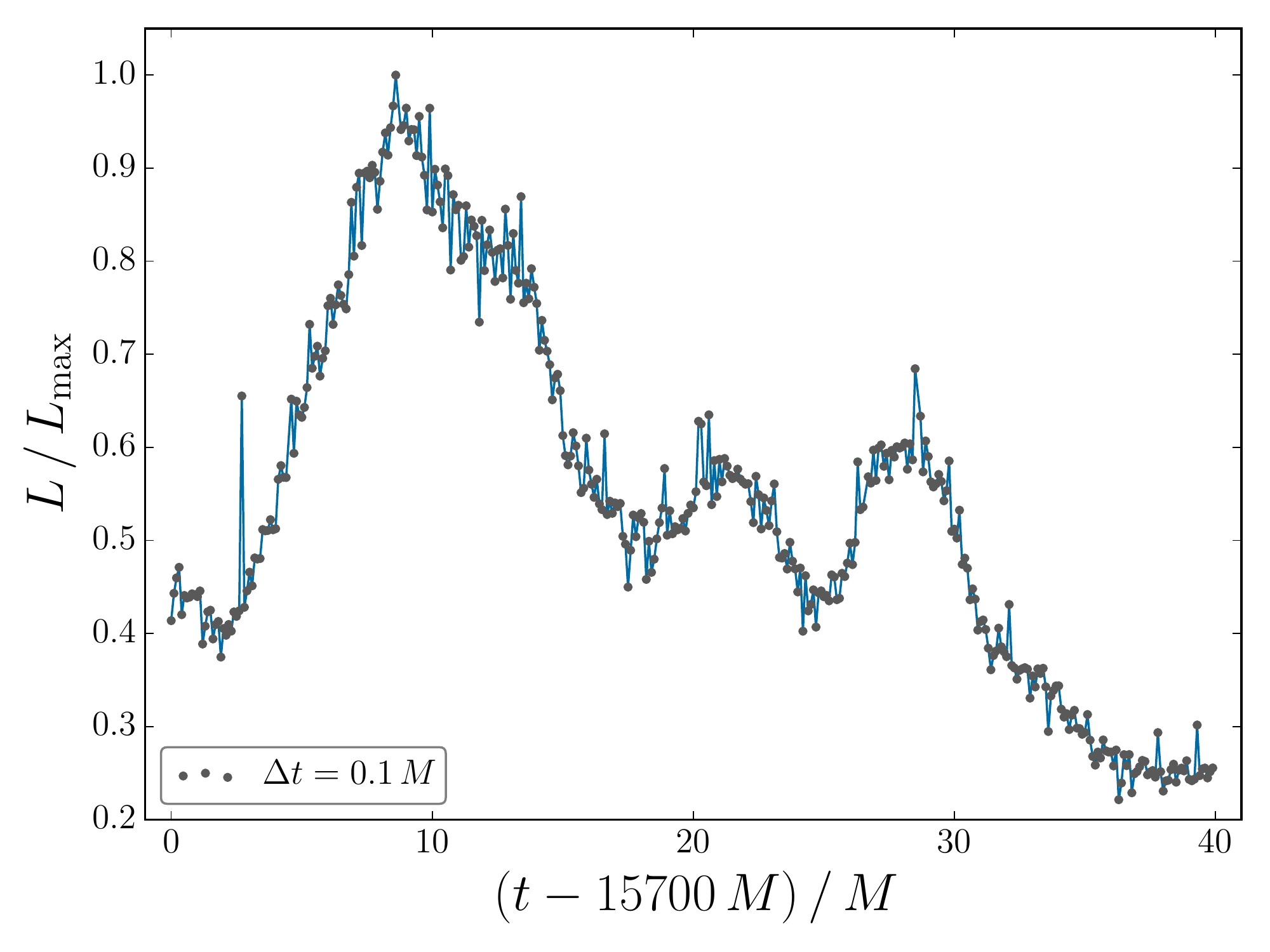}
    \caption{$\gamma$-ray lightcurve (frequency-integrated between 
        $10^{19}$ and $10^{24}$ Hz) with a high temporal resolution of 
        $\Delta t=0.1\,M$. 
        Although each time step corresponds to tracking 
        $\sim 10^{7}$ photons, variability on timescales shorter than 
        $\sim\text{few}\,\,M$ is significantly affected by noise due to 
        photon statistics.
        }
    \label{fig:lightcurve}
\end{figure}

The rough power law behaviour below $f\sim\,\text{few}\,\,\text{day}^{-1}$
changes to a cutoff at high frequencies.
The location of this cutoff is likely affected both by the spatial resolution of 
the fluid data, and the photon statistics of the radiative transport calculation.
Furthermore, the spectrum transitions to a flat white noise section above 
$10\,\,\text{day}^{-1}$. 
This is clearly due to poor photon statistics on the shortest timescales.
That is, the magnitude of the $\gamma$-ray variability is overwhelmed by
fluctuations due to photon statistics at timescales shorter than 
$\sim M$. In Figure~\ref{fig:lightcurve} we show one of the
$\gamma$-ray lightcurves with $\Delta t=0.1\,M$. Clearly, the variability on
short times is significantly affected by noise in the radiative
transport calculation.
The $\Delta t=0.1\,M$ lightcurve corresponds to our highest resolution 
simulations, tracking $\sim 10^{7}$ photons per time step, and so
improving upon this would be too computationally expensive.
In any case, our ``fast-light'' assumption limits the reliably of the radiative 
transport results on timescales shorter than $\sim \,M$.

\begin{figure}
    \centering
    \includegraphics[width=\figfactor\linewidth]{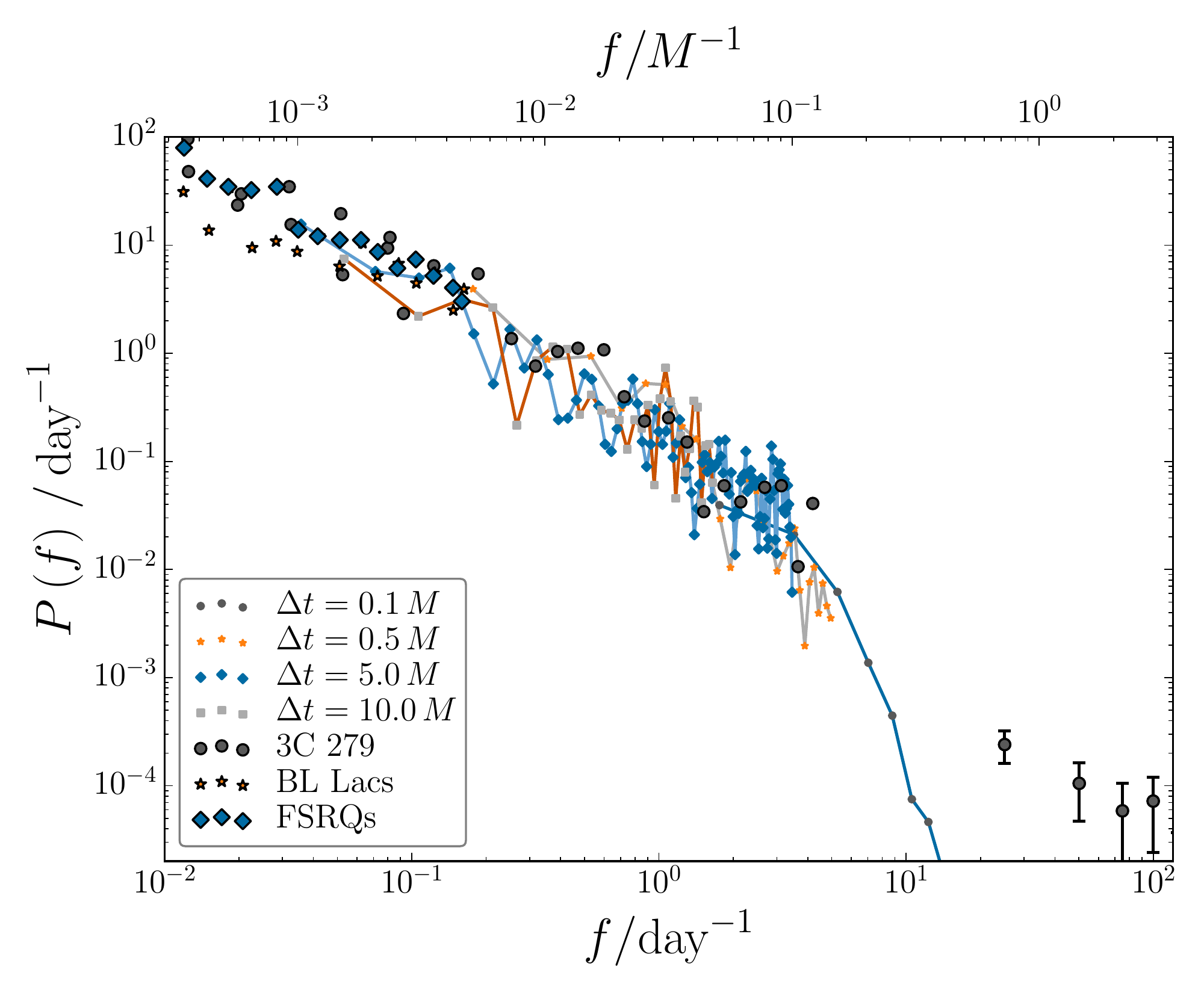}
    \caption{Same as Figure~\ref{fig:gamma_ray_psd} but showing the 
        simulated PDS for
        the optical band. 
        In this case, the lightcurves are dominated by 
        synchrotron emission. 
        The observational data again corresponds to energies $>100\,\text{MeV}$.
        It is clear that optical PDS shows a similar power-law 
        behaviour to that in the $\gamma$-rays.
}
    \label{fig:optical_psd}
\end{figure}

In Figure~\ref{fig:optical_psd} we show the PDS in the optical band. In this
case the radiation is primarily due to synchrotron emission. Interestingly, the 
variability in the optical band is the same as that in the $\gamma$-rays. 
Recently, \citet{Goyal+17} found that the $\gamma$-ray PDS in PKS 0735+178
is somewhat flatter than the corresponding PDS in the optical and radio bands.
They speculate that this discrepancy is due to additional stochastic processes 
which only affect the inverse Compton component. We will investigate this 
interesting observation in a future work.

\section{Summary and Discussion}
\label{sec:discussion}
As a step towards understanding the origin of variability
in blazars, we calculated the temporal dependence of optical and 
$\gamma$-ray radiation from the jet launching region in a MAD accretion flow.
In this work, we are concerned with two main properties of the observed 
variability.
Namely, (i) that the variability can be described by a PDS of power-law 
shape across the entire observed frequency range, and (ii) the very fast
variability observed in some sources, which can be significantly shorter than 
the light-crossing time of the supermassive black hole.
We argue that turbulence in the highly-magnetized plasma of a relativistic jet 
can plausibly account for both of these properties.

The radiation from our model is dominated by turbulent plasma close to the 
black hole.
The optical band is dominated by synchrotron emission, while the $\gamma$-rays 
are produced by inverse-Compton scattering from hot electrons.
Both the optical and $\gamma$-ray lightcurves show variability which can be 
characterized by a PDS with a power-law shape below 
$f\sim\,\text{few}\,\,\text{day}^{-1}$. 
Remarkably, the power-law section of our simulated PDS quantitatively 
reproduces the PDS observed in both BL Lacs and FSRQs.
This is suggestive of a connection between turbulent processes in MADs and
the observed variability properties of blazars.
Furthermore, we find the same power-law index for both the optical 
and $\gamma$-ray lightcurves, implying that the variability
properties should not be particularly sensitive to the underlying emission 
mechanism.

We also showed that the turbulent flow in the jet launching region naturally
produces structures on spatial scales much smaller than the horizon radius.
Such small scale structures, if produced at large distances in the 
highly-magnetized, relativistic jet, would likely produce significant
short timescale variability,
although it is indeed possible that the extremely fast 
variability in some sources might require additional microphysics to inject 
energy into the plasma on small scales \citep[e.g.,][]{GUB09,NP12}.
Based on these results, and assuming that the turbulence in the jet launching 
region persists to large radii, we argue that turbulence in the relativistic jet
is responsible for the observed variability in blazars over many orders
of magnitude in frequency.

While we have focussed on blazars, we expect the results presented here 
to also be applicable in a range of systems that can be modelled as
a radiatively inefficient accretion flow.
In particular, our results could be appropriate for Sgr A* and 
the low-hard state in XRBs, although observing variability in XRBs on 
timescales comparable to the light-crossing time of the black hole would be 
extremely challenging.
It would also be interesting to see how our results might change in the
super-Eddington MAD regime, which could be important for describing
ultra-luminous X-ray sources 
\citep{Sadowski+13,McKinney+14,McKinney+15,Sadowski+15b,Narayan+17}.

Our model has some limitations which prevent direct application in general, 
especially to systems with large luminosities.
Firstly, while our model might be applicable in its current form to some 
low-luminosity 
systems, the fact that the radiation originates from close to the black hole means 
that it is likely inappropriate for directly explaining the observed 
$\gamma$-rays from high-luminosity sources.
As discussed in Appendix~\ref{sec:pairs}, the $\gamma$-ray
radiation from high-luminosity systems (such as 3C 279) cannot
originate from close to the black hole since this region is too compact.
Such a compact region would be subject to a very large pair opacity and so the 
emitted
$\gamma$-rays would not escape to infinity. 
This implies that the $\gamma$-rays likely originate from a
region with a significant bulk Lorenz factor $\Gamma$ i.e., a relativistic jet.
If we assume that the intrinsic variability in the jet at large radii matches 
our results
for the inner jet launching region, then the observed variability should follow 
the same power-law, 
although shifted to higher frequencies by the Doppler factor
$\Dcal = \left(\Gamma\left(1-\beta\cos\theta\right)\right)^{-1}$.
The Lorentz boosting would also affect the overall normalization of the PDS, 
which we have not considered here since the scaling of the $y$-axes in
Figures~\ref{fig:gamma_ray_psd} and~\ref{fig:optical_psd} is arbitrary.

Secondly, the current work is limited by our assumption of a thermal 
distribution of electrons and by our simplified treatment of the electron 
thermodynamics. 
In reality, the electron distribution in highly-magnetized regions of the plasma 
likely has a high-energy non-thermal tail due to acceleration either by shock 
waves \citep[e.g.,][]{Sironi+15} or magnetic reconnection \citep[e.g.,][]{SS14}.
These non-thermal electrons would contribute
significantly to the observed radiation at very high energies.
Furthermore, the details of the electron physics in radiatively inefficient 
accretion flows \citep[RIAFs; e.g.,][]{NM08,YN14} remains a highly active area 
of research \citep{Ressler+15,Foucart+16,Sadowski+17a}. We will incorporate 
more complicated models of the electron physics in a future work.
Therefore, although our model reproduces the variability properties quite well,
the lack of non-thermal particles and large bulk Lorentz factor means that
we cannot quantitatively reproduce the wide range
of observed blazar spectra.

As mentioned previously, we have used a ``fast-light'' 
approximation in which the snapshots of the fluid data are treated as 
time-independent during the radiative transport calculation.
We will relax this approximation in a future work in order to study 
variability on shorter timescales in greater detail. 

We conclude by noting that a thorough investigation of the 
$\gamma$-ray zone in blazars will require 
more detailed microphysical modelling for treating the electrons, as well as 
global GRMHD simulations of self-consistent jet launching and propagation to 
large distances, with sub-$M$ grid resolution at large radii.
Recent MHD simulations investigated the large scale structure of
galactic jets propagating in an ambient medium \citep{TB16,Duran+16},
however, resolving small scale structures in both the inner accretion flow and 
the jet at large distances is currently too computationally expensive.


\acknowledgments 
The authors acknowledge the DJEI/DES/SFI/HEA Irish Centre for High-End Computing 
(ICHEC) for the provision of computational facilities under projects 
ucast007c and ucast008b.
MOR is supported by the Irish Research Council under grant number GOIPG/2013/315.
This research was partially supported by the European Union Seventh Framework 
Programme (FP7/2007-2013) under grant agreement no 618499.
JCM acknowledges NASA/NSF/TCAN (NNX14AB46G), NSF/XSEDE/TACC (TGPHY120005), and 
NASA/Pleiades (SMD-14-5451).
The authors would like to thank the anonymous referee for helpful and 
constructive comments.
\software{\texttt{HARM} \citep{Gammie+03}, \texttt{grmonty} \citep{Dolence+09}, SciPy \citep{scipy}}

\appendix

\section{Pair Opacity}
\label{sec:pairs}
Here, we estimate constraints on the luminosity from the pair opacity.
The following applies in the comoving frame of the source and follows closely 
the treatment of \citet{DG95}.
High-energy $\gamma$-rays can collide with softer
radiation to produce $e^{\pm}$ pairs. The cross section for this
process is maximized for collisions between $\gamma$-rays of energy
$x=h\nu/mc^{2}$ and target photons of energy $x_{t}=1/x$. This maximum cross 
section is $\sigma = 3\sigma_{T}/16$ \citep[e.g.,][]{Lang80}, 
where $\sigma_T$ is the Thomson cross section. 
Assuming that the source is spherical and emits isotropically, 
the corresponding optical depth can be written in terms of the luminosity as
\begin{equation}
\tau_{\gamma\gamma}\left(x\right)=\frac{3\sigma_{T}}{16}\frac{L\left(x_{t}\right)}{4\pi mc^{3}R}
\end{equation}
Therefore, the condition that $\tau_{\gamma\gamma}\left(x\right)\lesssim 1$
constrains the luminosity of the soft radiation field to be
\begin{equation}
L\left(x_{t}\right)\lesssim 
3\times10^{43}\,M_8\,\left(\frac{R}{r_{g}}\right)
\text{erg s}^{-1}
\end{equation}
where we have written the luminosity in terms of the gravitational radius
$r_{g}\approx1.5\times10^{13}\,M_8\,\text{cm}$, and $M_8$ is the black hole mass
in units of $10^8\,M_\odot$.

In the case of 3C 279, the observed $\gamma$-rays primarily interact
with a soft (X-ray) radiation field of luminosity 
$L\sim 3\times 10^{47}\,\text{erg s}^{-1}$ \citep{Ackermann+16}.
Adopting a black hole mass of $M=5\times 10^{8}\,M_\odot$ and an emission region
of size $R=5\,r_g$, we find that the luminosity of the target radiation is 
constrained to be below $L\lesssim 10^{45}\,\text{erg s}^{-1}$.
The obsereved luminosity is more than two orders of magnitude larger than this.
Combined with the short variability timescale, we are forced to
to conclude that the source of emission is strongly Doppler 
boosted and so likely originates in a relativistic jet.



\bibliographystyle{hapj}
\bibliography{variability}

\end{document}